\documentclass[twocolumn,aps,floatfix]{revtex4}
\usepackage{amssymb} \usepackage{graphicx} \usepackage{amsmath}
\usepackage[T1]{fontenc} \usepackage{pstricks} \usepackage{subfigure}
\usepackage[colorlinks=true,citecolor=blue,urlcolor=red]{hyperref}
\usepackage[normalem]{ulem}

\newcommand{\mbf}[1]{\mathbf{#1}}

\newcommand{\av}[1]{\left\langle #1 \right\rangle}

\begin{document}
\title{Signatures of a universal jump in the superfluid density in two-dimensional Bose gas with finite number of particles }

\author{Krzysztof Gawryluk$^{1}$ and Miros{\l}aw Brewczyk$\,^{1}$}

\affiliation{\mbox{$^1$Wydzia{\l} Fizyki, Uniwersytet w
    Bia{\l}ymstoku,  ul. K. Cio{\l}kowskiego 1L, 15-245 Bia{\l}ystok,
    Poland}   }

\date{\today}

\begin{abstract}
We study, within the classical fields approximation, a two-dimensional weakly interacting uniform Bose gas of a finite number of atoms. By using a grand canonical ensemble formalism we show that such systems exhibit, in addition to the Berezinskii-Kosterlitz-Thouless (BKT) and thermal phases, an intermediate region. This intermediate region is characterized by a decay of current-current correlations at low momenta and by an algebraic decay of the first-order correlations with an exponent being larger than the critical value predicted by the BKT theory. The density of the superfluid fraction at the temperature which separates the BKT phase from the intermediate region approaches the one found by Nelson and Kosterlitz for two-dimensional superfluids while the number of atoms is increased.

\end{abstract}

\maketitle

One- and two-dimensional systems possess unusual properties. For instance, thermal fluctuations unable the phase transition to Bose-Einstein condensate in low dimensional Bose gases \cite{Hohenberg}. Instead, a two-dimensional Bose systems exhibit a new kind of phase transitions, related to spontaneous creation of vortices \cite{Berezinskii71,Kosterlitz72}. Below the BKT transition temperature vortices form tight pairs, whereas above the pairs break and the vortices move on their own. In a one-dimensional Bose gas, on the other hand, different kind of excitations are developed, which were studied theoretically by Lieb and Liniger \cite{LiebLiniger} and whose existence was recently confirmed by the classical fields calculations \cite{Karpiuk1215}. These excitations - spontaneous solitons - are analogous to pairs of vortices in a two-dimensional case.

As shown by Nelson and Kosterlitz \cite{Nelson77}, a two-dimensional superfluid exhibits a jump of the density at the critical temperature. Theory developed in \cite{Kosterlitz73,Kosterlitz74} predicts that the ratio of the superfluid density at the critical temperature to the critical temperature depends only on the fundamental constants and equals $2\, m^2 k_B/\hbar^2 \pi$. This relation has been confirmed experimentally with different physical systems, including thin superfluid $^4$He films adsorbed on a solid substrate \cite{Bishop78}, thin films of superconductors \cite{Hebard80,Epstein81}, and planar arrays of Josephson junctions \cite{Resnick81}.

All these experiments brought evidences supporting the presence of the BKT phase transition in a two-dimensional systems, however, none of them proved directly the existence of underlying mechanism of binding and unbinding pairs of vortices. Gaining direct evidences for that became possible only in the era of cold atoms. Experimental studies of a two-dimensional Bose gas began soon after the first achievement of the Bose-Einstein condensate \cite{Hadzibabic06,Gorlitz01,Orzel01,Burger02,Schweikhard04,Rychtarik04,Hadzibabic04,Smith05,Kohl05}. One of the first experimental approach to the BKT phenomenon in a gas of cold atoms was reported in Ref. \cite{Hadzibabic06}. It was shown that the BKT description applies to the finite-size systems although the transition resembles the crossover rather than the sharp phase transition. The other important result was related to an observation of vortex proliferation which began abruptly while the temperature was increasing.

Most of the ultracold-atom experiments on BKT phase to date were performed with a gas confined in a two-dimensional harmonic trap \cite{Hadzibabic06,Schweikhard07,Kruger07,Clade09,Tung10,Rath10,Plisson11,Hung11,Yefsah11,Desbuquois12,Choi12,Ha13,Choi13,Desbuquois14,Ries15,Fletcher15}. This introduces a new degree of freedom into play since the Bose-Einstein condensation (BEC) phase transition becomes possible. An interplay between the interaction-driven BKT phase transition and the Bose-Einstein condensation has been experimentally studied in \cite{Fletcher15}. The emergence of coherence in a sample with tunable interaction was observed and attributed to the BKT superfluid transition. It was shown that the BKT transition converges to the BEC one when the interactions are vanishing. With appearance of a possibility of trapping atoms in a uniform potential \cite{Gaunt13,Corman14} new studies of emergence of coherence in a two-dimensional Bose gas confined in a box-like potential has started \cite{Chomaz15}. Here, we numerically investigate the properties of BKT phase in a uniform two-dimensional Bose gas with finite number of atoms, in particular, we calculate the superfluid density in the sample. Although such quantity could be identified by the measurement of the speed of second sound (see Ref. \cite{Ota18a} and a recent experimental work on sound propagation in two-dimensional Bose gas \cite{Ville18} as well as its theoretical description \cite{Ota18b}), we propose here to investigate the current-current correlations.

Numerical studies of the BKT transition in a two-dimensional Bose gas requires the knowledge of techniques treating nonzero temperatures. Such methods have been already well developed for degenerate Bose gases \cite{review, Proukakis, Blakie}. In particular, $c$-field methods as described in Ref. \cite{Blakie} were used to thoroughly discuss the properties of both trapped \cite{Simula06,Simula08,Bisset09} and uniform \cite{Foster10} two-dimensional Bose gases. Here, we are following the classical fields approximation \cite{review}. We are interested in equilibrium states, hence we built the statistical ensemble of classical fields \cite{Witkowska10,Gawryluk17,Pietraszewicz17}. In the present paper the grand canonical ensemble is used, therefore the temperature and the chemical potential are two control parameters. They determine uniquely the average total number of atoms in the system, which is in our case of the order of a few thousand. We change the temperature in a wide range to find the transition to the thermal phase which is characterized by the exponential decay of the first-order correlation function as well as by the extinction of the superfluid fraction. Increasing the temperature moves the energy cutoff, typical for $c$-field methods \cite{Zawitkowski04,Witkowska09,Bienias11,Pietraszewicz15}, up forcing the usage of larger sets of basis functions needed for the expansion of the classical field.

The superfluid density of a two-dimensional uniform Bose gas of $N$ particles occupying a volume $V$ can be obtained by calculating the current-current correlations in momentum space, derived based on the hydrodynamic theory of superfluid \cite{PitaevskiiStringari} (for an equivalent approach utilizing the momentum density correlations, see Refs. \cite{Foster10} and \cite{Forster})  
\begin{equation}
\av{(j_{\mbf{k}})_{\ell}(j_{\mbf{k}})_m^{\star}}=\frac{\varrho_s}{m^2}\frac{k_B T}{k^2}k_{\ell}k_m V+\frac{\varrho_n}{m^2}k_BT \delta_{\ell m}V  \,.
\label{cccor}
\end{equation}
The above formula is valid in the limit of vanishing momentum. The current density, $\mbf{j}(\mbf{r})$, itself is initially determined in coordinate space as
\begin{equation}
\mbf{j}(\mbf{r})=\frac{i \hbar}{2 m}\Big((\nabla \psi^{\star}(\mbf{r})) \psi(\mbf{r})-\psi(\mbf{r})^{\star} \nabla \psi(\mbf{r})\Big)  \,,
\label{current}
\end{equation}
where $\psi(\mbf{r})$ is the classical field, and then transformed to momentum space. To get the superfluid, $\varrho_s$, and normal, $\varrho_n$, densities one needs to utilize Eq. (\ref{cccor}). First, considering the $xy$ (i.e., when $l=x$ and $m=y$) correlations
\begin{equation}
\av{(j_{\mbf{k}})_{x}(j_{\mbf{k}})_y^{\star}}=\frac{\varrho_s}{m^2} \frac{k_B T}{k^2} k_x k_y V
\label{corxy}
\end{equation}
allows to determine the superfluid density. Then, by calculating the $xx$ (i.e., when $l=m=x$) correlations we can easily deduce the density of normal fraction
\begin{equation}
\av{(j_{\mbf{k}})_{x}(j_{\mbf{k}})_x^{\star}}=\frac{\varrho_s}{m^2} \frac{k_B T}{k^2} k_x k_x V
+\frac{\varrho_n}{m^2}k_BT \, V  \,.
\label{corxx}
\end{equation}

\begin{figure}[bth] 
\includegraphics[width=5.9cm]{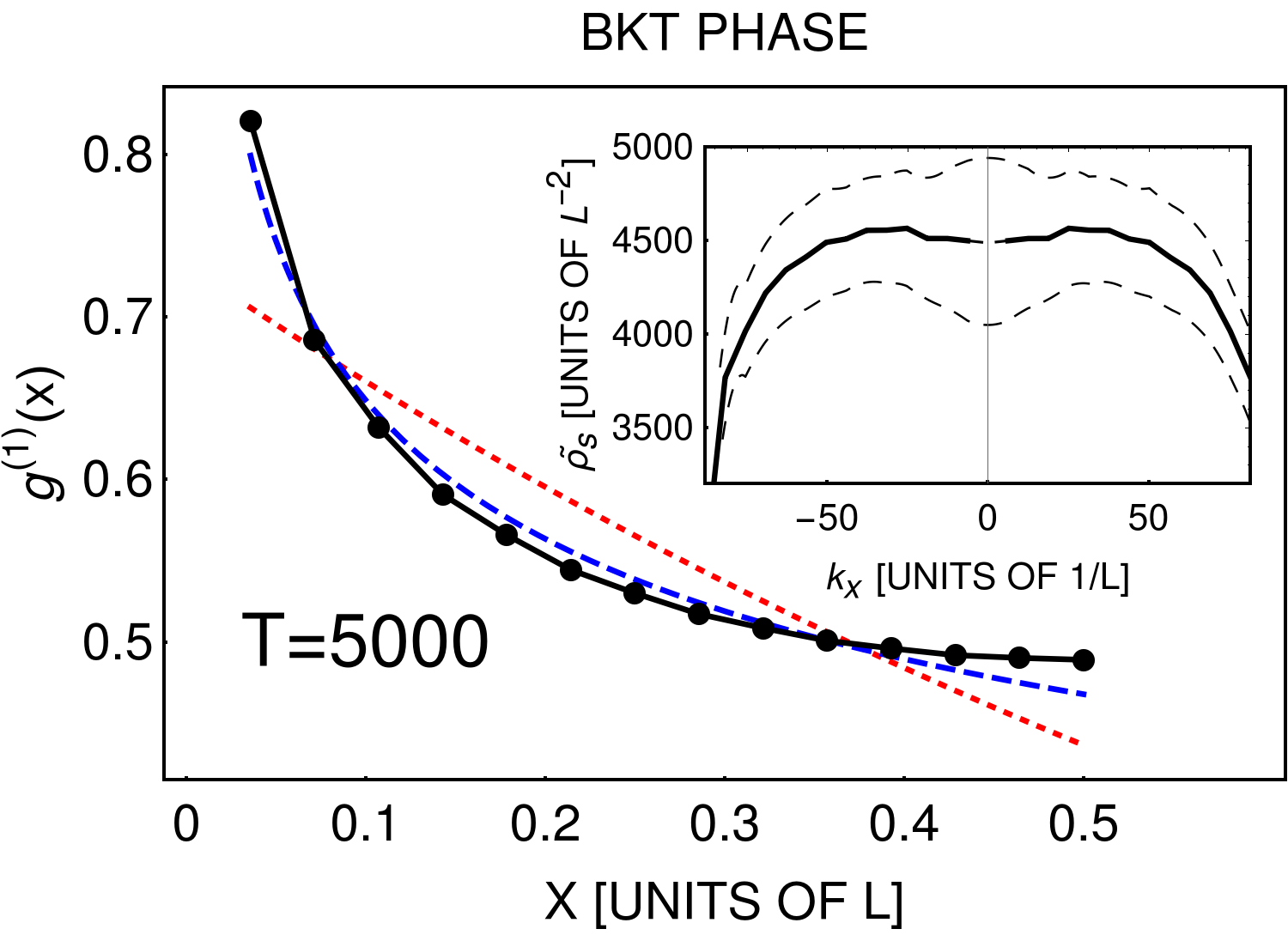}  \\   \vspace{0.4cm}
\includegraphics[width=5.9cm]{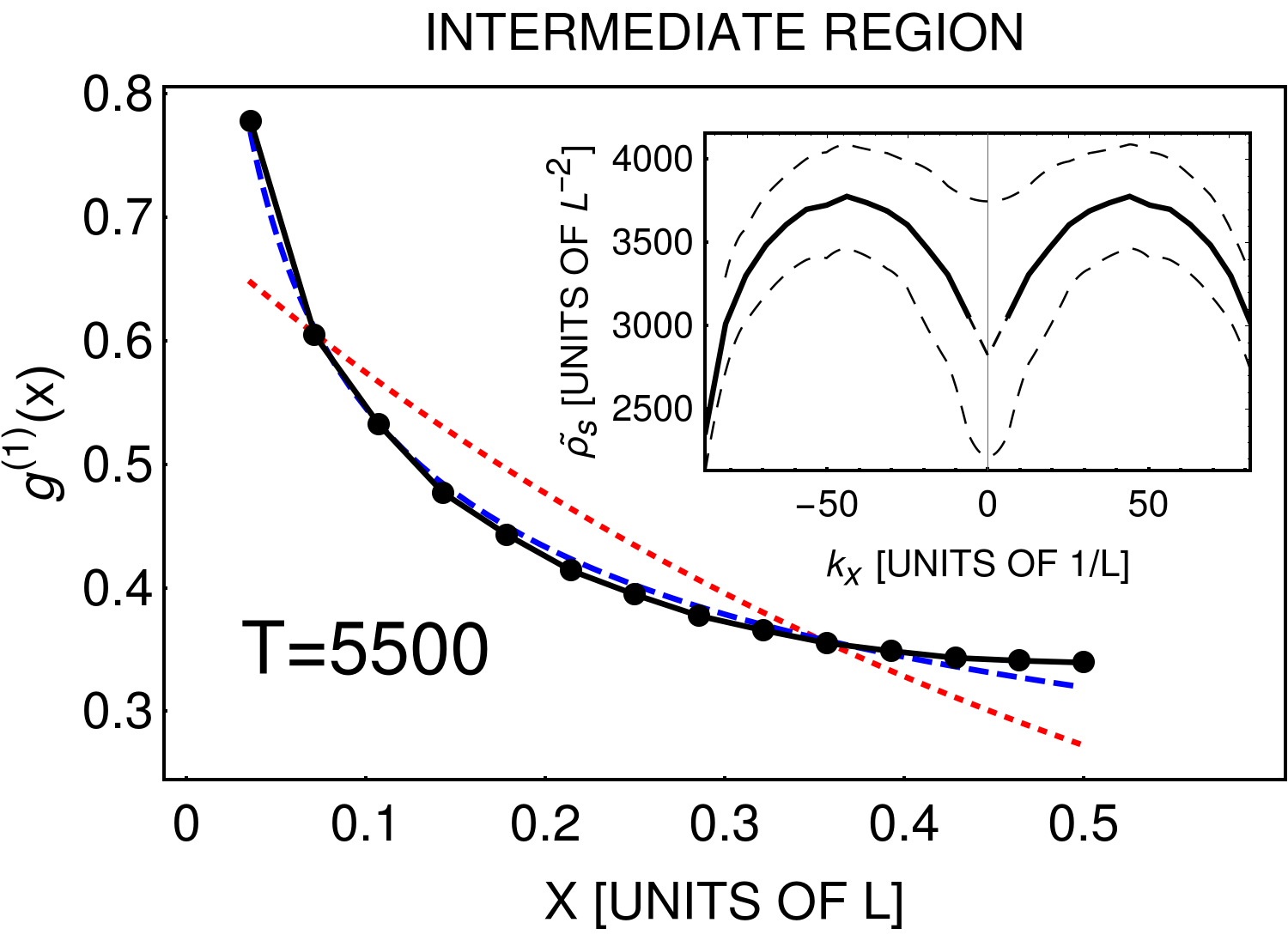}   \\   \vspace{0.4cm}
\includegraphics[width=5.9cm]{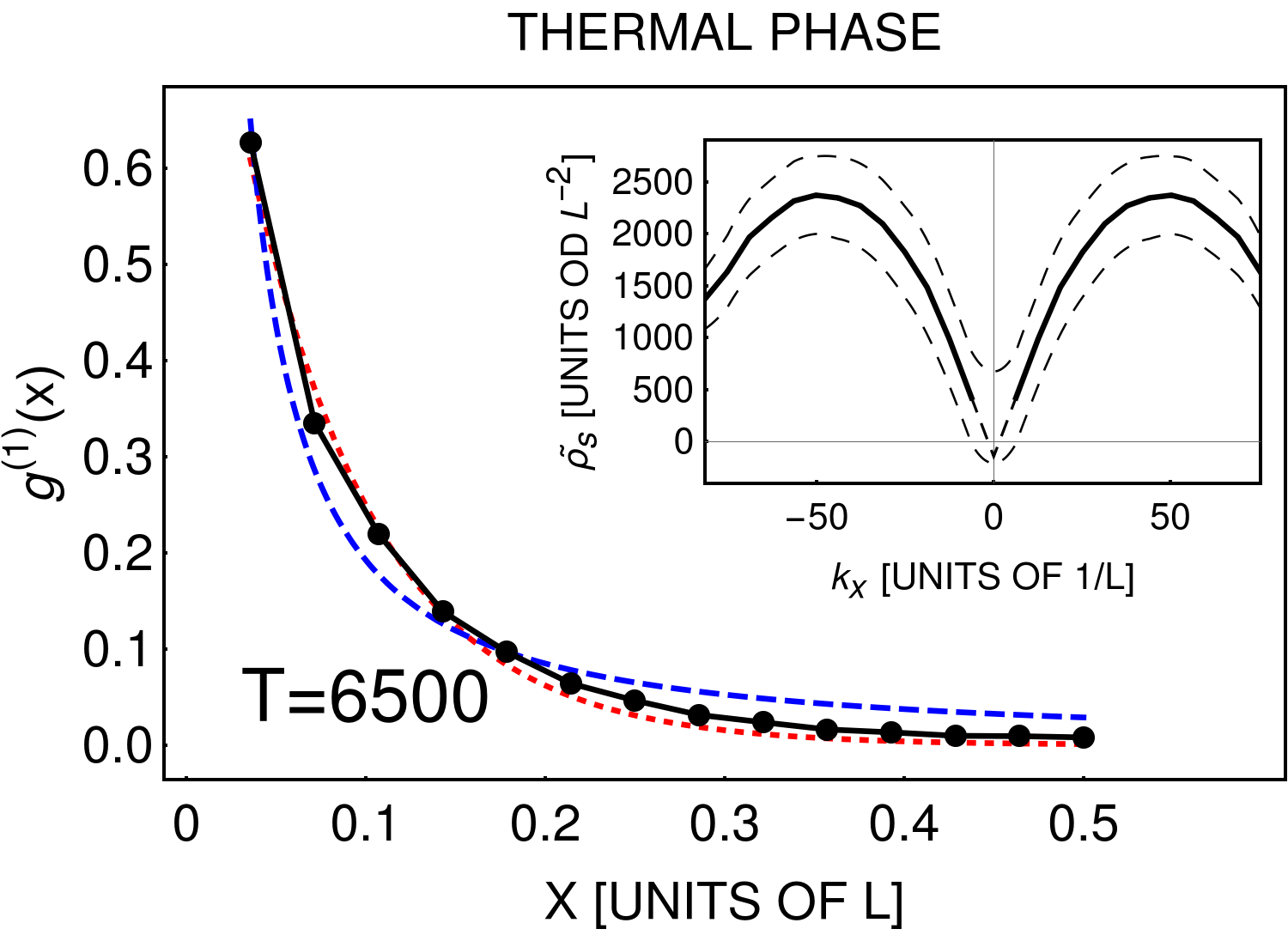}
\caption{First-order correlation function, $g^{(1)}(x)$, representative for the BKT phase (upper frame), the intermediate region (middle frame), and the thermal phase (lower frame). Black solid lines are the outcome of our numerical procedure, blue dashed lines stand for the best algebraic fits, whereas the red dotted curves represent the best exponential fits. Evidently, for the BKT and intermediate phases the algebraic decay behavior works better. However, for the case of intermediate region the exponent is larger than the critical value given by $0.25$ what is the characteristics of this region. Above the transition temperature the exponential decay of $g^{(1)}(x)$ becomes clear (lower frame). Insets (solid lines) show the behavior of the current-current correlations in the momentum space. The onset of the intermediate region is defined by the temperature at which the current-current correlations start to diminish at low momenta. The dashed lines represent the standard deviation of $(j_{\mbf{k}})_{x}(j_{\mbf{k}})_y^{\star}$. For all frames $g=0.35\, \hbar^2/m$.  }  
\label{g1}
\end{figure}

We do calculate the superfluid fraction of a two-dimensional uniform Bose gas according to the above prescription. For a given chemical potential we scan the temperature and for each temperature we find the $xy$ and $xx$ current-current correlations as a function of momentum in two-dimensional space. We average the $(j_{\mbf{k}})_{\ell}(j_{\mbf{k}})_m^{\star}$ product over the grand canonical ensemble. To make the outcome smooth enough we additionally do averaging over time while propagating each classical field according to the Gross-Pitaevskii equation \cite{review}. The solution of Eq. (\ref{corxy}), which we denote here as $\tilde{\varrho_s}$, is shown in Fig. \ref{g1} in the insets along the line $\mbf{k}=(k/\sqrt{2},k/\sqrt{2})$. The superfluid density, $\varrho_s$, is obtained from $\tilde{\varrho_s}$ by calculating the limit of zero momentum. The Eq. (\ref{corxx}) gives us the normal component density, hence an internal consistency of our numerical procedure can be checked -- the total density must equal the one obtained within the ground canonical ensemble approach.

\begin{figure}[thb] 
\includegraphics[width=7.4cm]{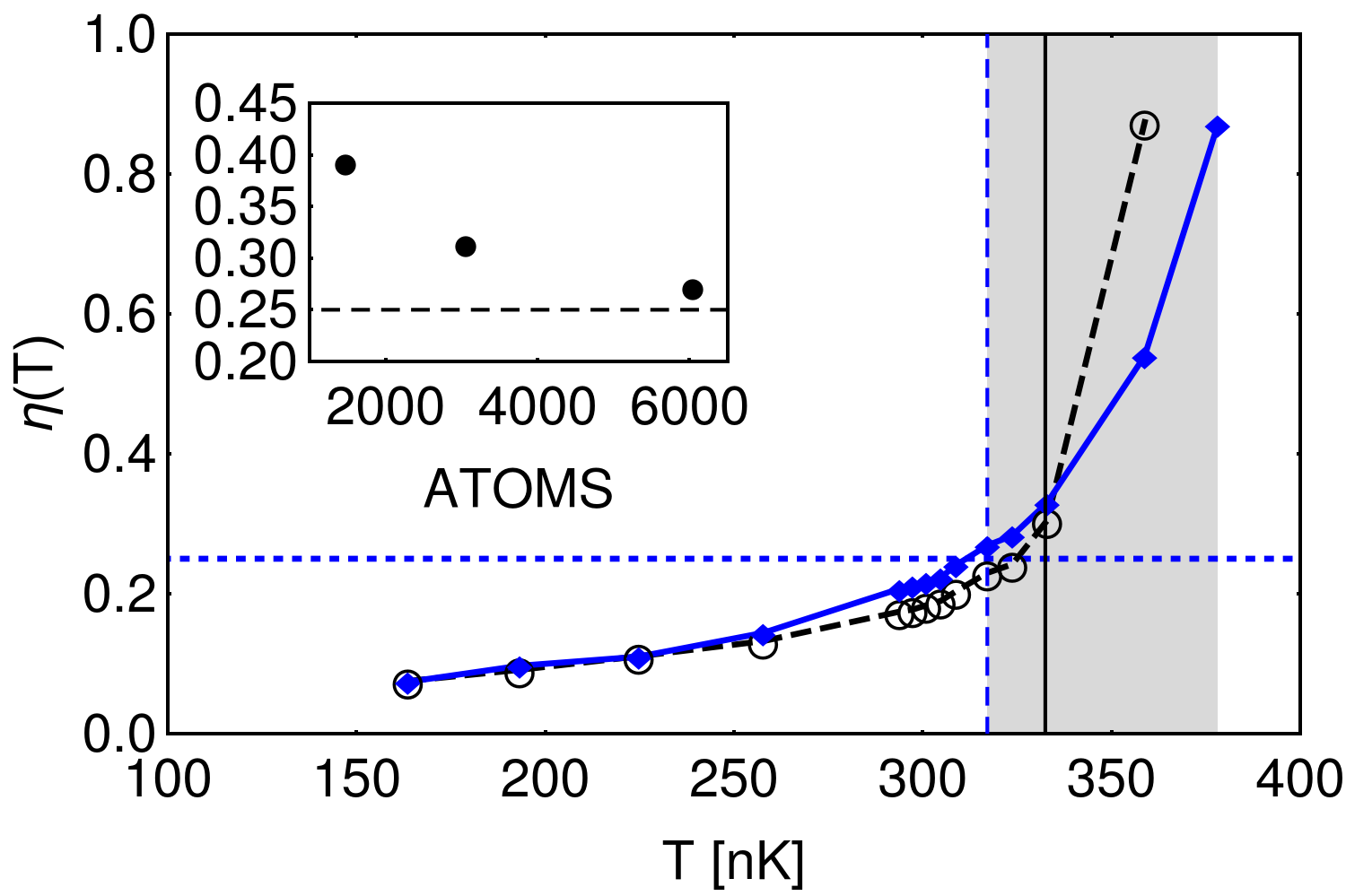} 
\caption{Main frame: Exponent $\eta(T)$ (solid line) determined by fitting the first-order correlation function obtained numerically to the algebraic fall off formula as a function of temperature. The solid line shows the results for $\mu \approx 2500$ ($N\approx 6000$ and atomic density of $66.2\,$$\mu m^{-2}$) and $g=0.35$. The dashed line comes from the formula $\eta(T)=m^2 k_B T/2\pi \hbar^2 \varrho_s(T)\,$ \cite{Nelson77}, with the superfluid density obtained numerically. Vertical dashed line is the transition temperature, $T_{tr}$, to the intermediate region (marked by a shaded area) whereas the solid one shows the transition to the BKT phase in the thermodynamic limit \cite{Prokofev01}. Horizontal dashed line is the value of the critical exponent. Inset: Exponent $\eta(T)$ at the transition temperature, $T=T_{tr}$, to the intermediate region as a function of number of atoms. Errors in the main panel and in inset are smaller than the size of symbols used. Clearly, $\eta(T_{tr})$ approaches the critical value of $0.25\,$ \cite{Nelson77} with increasing number of atoms.   }  
\label{exptemp}
\end{figure}

Fig. \ref{g1} includes three frames, which are representative for the BKT phase (upper frame), the intermediate region (defined below, middle frame), and the thermal phase (lower frame). Here, the interaction strength $g=0.35\, \hbar^2/m$ and the chemical potential $\mu = 2500\, \hbar^2/m L^2$ ($L$ is the length of a two-dimensional box). The number of atoms for each system is about $N=6000$. The upper frame is typical for the BKT phase, where the current-current correlations are flat for small momenta but their value changes with temperature. Here, the temperature is $5000\, \hbar^2/ m L^2 k_B$. When the temperature increases, this behavior gets modified qualitatively. Above some characteristic temperature, $T_{tr}$, the current-current correlations start to diminish for small momenta. We say that the system enters the intermediate regime (middle frame in Fig. \ref{g1}). In this region the superfluid density rapidly goes to zero. When the superfluid density vanishes the system reaches the thermal phase (lower frame in Fig. \ref{g1}).

This behavior of the current-current correlations coordinates with the properties of the first-order correlation function, $g^{(1)}(\mbf{r},\mbf{r'})$, defined as a normalized average $g^{(1)}(\mbf{r},\mbf{r'})=\av{\psi^{\star}(\mbf{r})\, \psi(\mbf{r'})} /\av{\psi^{\star}(\mbf{r})} \av{\psi(\mbf{r'})}$ over the grand canonical ensemble (in order to improve the quality of results, additional averaging over time while propagating the classical field, is applied). Main frames in Fig. \ref{g1} (black solid lines) show the $g^{(1)}(\mbf{r},0)$ function for all characteristic regimes. For temperatures below the BKT transition temperature the system exhibits the quasi-long-range order, i.e. $g^{(1)}(\mbf{r},0)$ decays algebraically with a distance, $g^{(1)} \propto r^{-\eta}$ (as opposed to the exponential decay occurring for the thermal gas), with a temperature dependent exponent $\eta(T)$ (see Fig. \ref{exptemp}). As shown by Nelson and Kosterlitz, for an infinite system this exponent is related to the superfluid density by $\eta(T)=m^2 k_B T/2\pi \hbar^2 \varrho_s(T)$ \cite{Nelson77}. The critical value of the exponent, i.e. its value at the critical temperature, equals $0.25$. Indeed, our results prove the fall off of the correlations with a distance according to the power law (see Fig. \ref{g1}, upper frame), with the exponent increasing with temperature (see Fig. \ref{exptemp}). Note that all data available for $g^{(1)}(\mbf{r},0)$ are used to get the best algebraic fit. In fact, there is some discrepancy visible for large distances. This happens because the system we consider consists of finite number of atoms. This results in macroscopic occupation of zero momentum mode. Therefore, the correlations saturate at large distances to the value equal to the fraction of atoms being in the zero momentum state.

Within the intermediate region (shaded area in Fig. \ref{exptemp}, $T>T_{tr}$) the $g^{(1)}(\mbf{r},0)$ function still fits to the algebraic decay, not to the exponential one (we applied $\chi^2$ test to judge that, see Appendix \ref{first}) but now the exponent $\eta(T)>0.25$ (exponents larger than $1/4$ were also reported in \cite{Foster10} for ultracold atomic Bose gases and in \cite{Roumpos12,Dagvadorj15} for polariton systems). When the temperature gets beyond the one characterizing the intermediate region, the $g^{(1)}$ function changes qualitatively its character and starts to fit the exponential function (lower frame in Fig. \ref{g1}) -- also the superfluid fraction vanishes.

\begin{figure}[thb] 
\includegraphics[width=6.8cm]{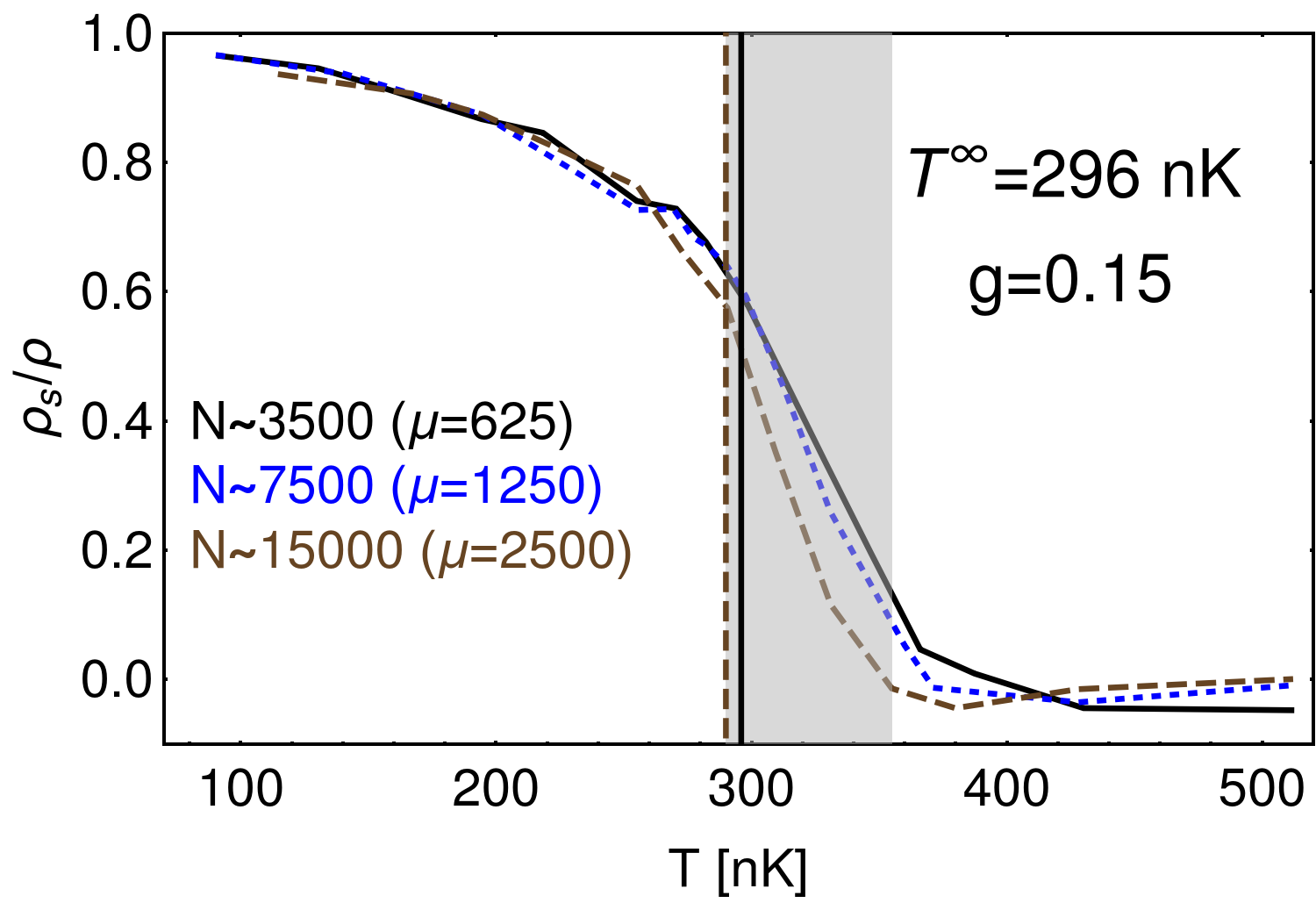}  \\   \vspace{0.4cm}
\includegraphics[width=6.8cm]{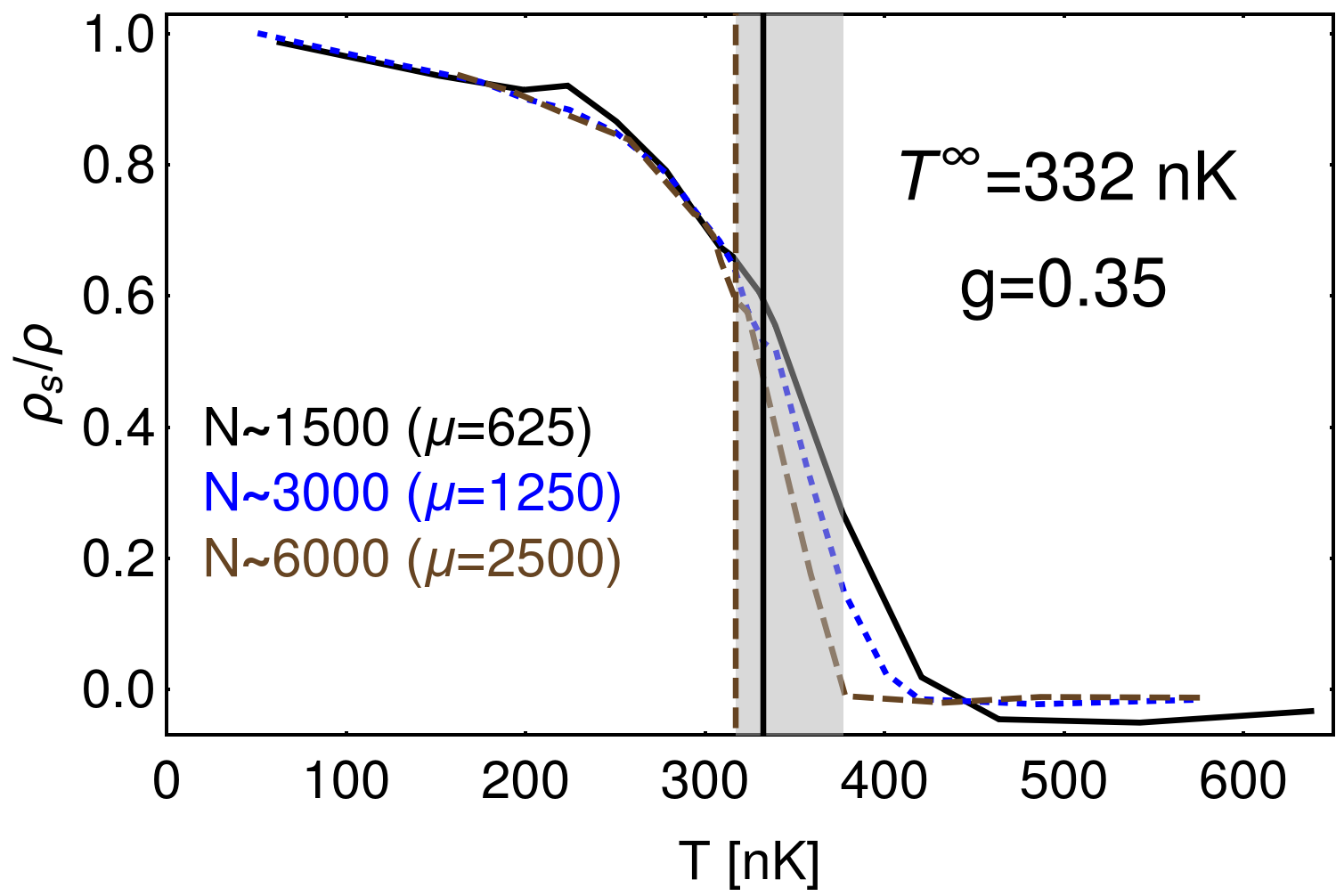}
\caption{Superfluid fraction as a function of temperature for a two-dimensional weakly interacting Bose gas with the interaction strength $g=0.15\, \hbar^2/m$ (upper frame) and $g=0.35\, \hbar^2/m$ (lower frame). There are three sets of curves in each frame corresponding to different number of atoms in the system. Going from the right to the left (from the top to bottom in the legend) the number of atoms increases as depicted in figures. Clearly, the intermediate region shrinks when the number of atoms grows sustaining the atomic density (here, equal to $66.2\,$$\mu m^{-2}$), i.e the system moves towards the thermodynamic limit. The shaded areas depict the intermediate regions for the highest number of atoms cases. The vertical solid black lines show the critical temperature for the infinite system calculated from the formula $T^{\infty} = 2\pi \hbar^2 n / m k_B \ln{(\xi \hbar^2/mg)}$ \cite{Prokofev01}, whereas the dashed color ones represent the transition, $T_{tr}$, temperatures found numerically and corresponding to the case of maximal number of atoms considered. Note that both temperatures, obtained by different approaches, are already very close to each other.} Also note that for larger number of atoms the curves become steeper.     
\label{superfrac}
\end{figure}

Fig. \ref{exptemp} shows the exponent $\eta(T)$ for the interacting ($g=0.35$) system consisting of $N=6000$ atoms. The numerical results almost follow the formula $\eta(T)=m^2 k_B T/2\pi \hbar^2 \varrho_s(T)$ \cite{Nelson77}, valid in the thermodynamic limit. The discrepancy appears close to the transition temperature, $T_{tr}$, to the intermediate regime (shaded area in the figure). This is because the system under consideration is finite and the value of the superfluid density, $\varrho_s(T)$, appearing in the expression for $\eta(T)$ is overestimated as taken for the system with finite number of atoms. It is emphasized also by the fact that the transition temperature to the BKT phase in the thermodynamic limit, given by $T^{\infty} = 2\pi \hbar^2 n / m k_B \ln{(\xi \hbar^2/mg)}$ with $\xi \approx 380$ (vertical solid line) \cite{Prokofev01}, is shifted with respect to $T_{tr}$. In the inset of Fig. \ref{exptemp} we collect the values of $\eta(T)$ at the transition temperatures, $T_{tr}$, for the gas with different number of atoms. Clearly, moving towards the infinite system gets the value of $\eta(T)$ closer to the value of the critical exponent, i.e. $0.25$.

\begin{figure}[t!bh] 
\includegraphics[width=8.2cm]{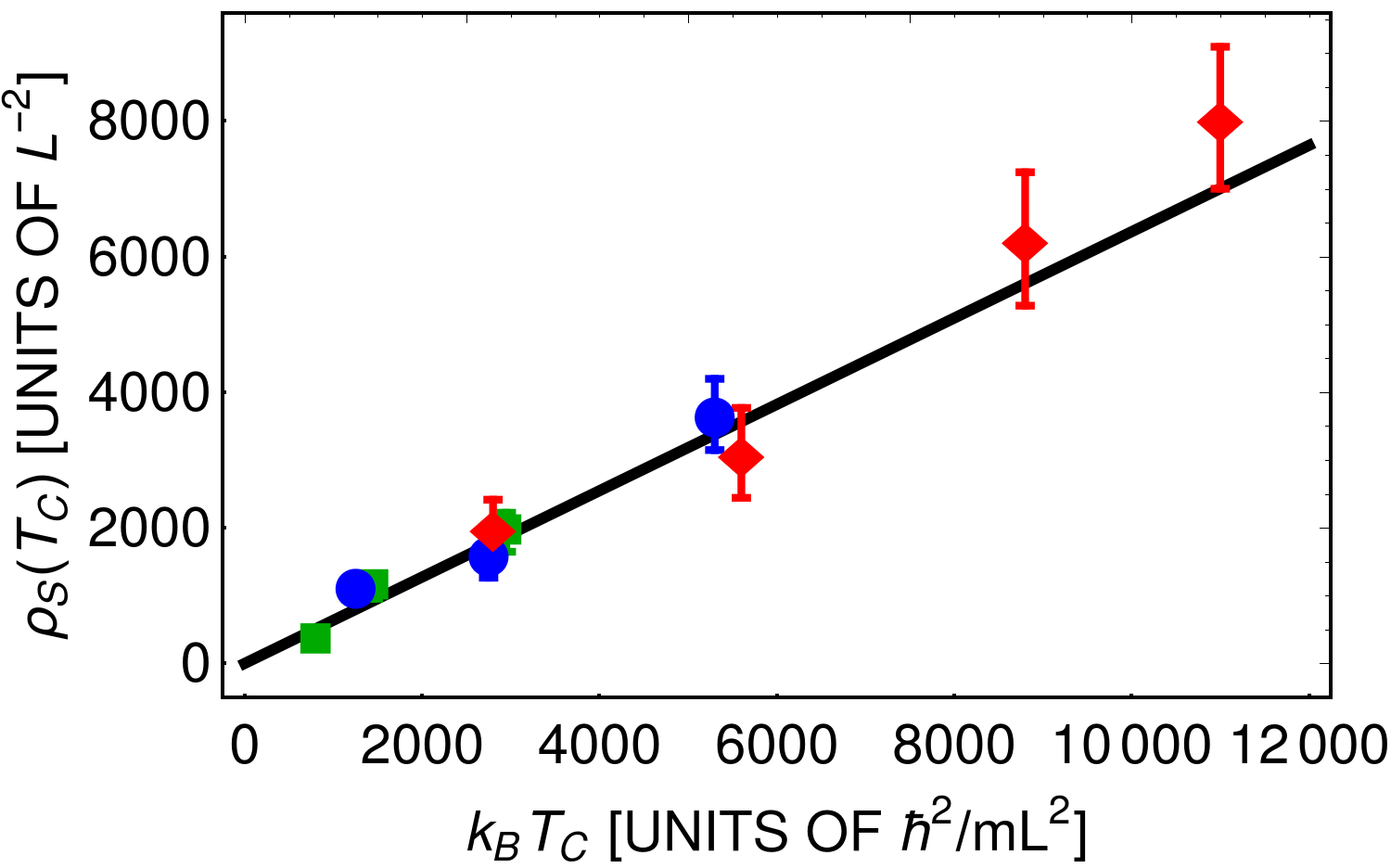} 
\caption{Superfluid fraction, $\rho_s(T_{tr})$, at the transition temperature $T_{tr}$ (symbols). The transition temperature is determined from the criterion based on the behavior of the current-current correlations at low momenta. For a given interaction strength ($g=0.7$ -- green squares, $g=0.35$ -- blue circles, and $g=0.15$ -- red diamonds), data for the system with different numbers of atoms (as in Fig. \ref{superfrac}) are shown. The solid line is the linear relation between the superfluid density at the BKT transition and the BKT transition temperature, $\rho_s(T_c) = (m^2 k_B/\hbar^2)\, 2\, T_c/\pi$, as proved by Nelson and Kosterlitz \cite{Nelson77} and experimentally verified in \cite{Bishop78,Hebard80,Epstein81,Resnick81}. As argued in the text, the transition temperature, $T_{tr}$, approaches the critical one when the number of atoms is increased. In the figure both temperatures are put equal.   }  
\label{universal}
\end{figure}

In Fig. \ref{superfrac} we demonstrate how the two-dimensional Bose gas behaves while approaching the thermodynamic limit. Here, we plot the superfluid fraction as a function of temperature in the system having a constant density (equal to $66.2\,$$\mu m^{-2}$) but increasing number of atoms. When the number of atoms gets larger the intermediate region shrinks and the transition temperature, $T_{tr}$, approaches the critical temperature (vertical solid black lines) \cite{Prokofev01}. At the same time the superfluid fraction decreases. Evidently, the transition temperature, $T_{tr}$, acquires the meaning of the critical temperature of a two-dimensional Bose gas when the number of atoms gets larger.

In Fig. \ref{universal} we summarize our results showing the superfluid density at the transition temperature $T_{tr}$. It is clear that the ratio $\rho_s(T_{tr})/T_{tr}$ follows the universal behavior $\rho_s(T_c) /T_c = (m^2 k_B/\hbar^2)\, 2/\pi$ proved in \cite{Nelson77}, assuming $T_{tr}=T_c$. Hence, a two-dimensional weakly interacting Bose gas belongs to the class of systems possessing properties (like the critical exponent for the decaying first-order correlations or the universal jump of the superfluid density at the transition) well understood within the BKT theory.

In summary, we have studied a weakly interacting two-dimensional Bose gas at thermal equilibrium, consisting of a finite number of atoms. In addition to the BKT and thermal phases, we identify the intermediate region. It is characterized by an algebraic decay of the first-order correlation function, as the BKT phase is, but with the decay exponent larger than the critical value and, simultaneously, by decrease of the current-current correlations for low momenta. When the number of atoms increases the intermediate region shrinks and the temperature separating the BKT and intermediate phases approaches the critical temperature. At the same time, the superfluid density at the transition temperature becomes the density which represents the universal jump of the superfluid density characteristic for two-dimensional systems discussed by Nelson and Kosterlitz \cite{Nelson77}.

\acknowledgments  
We are grateful to P. Deuar, M. Gajda, and K. Rz\c{a}\.zewski for helpful discussions. Part of the results were obtained using computers at the Computer Center of University of Bialystok.

\appendix
\section{First order correlation function}
\label{first}

\begin{figure*}[h!tb]
\centering
\includegraphics[width=5.5cm]{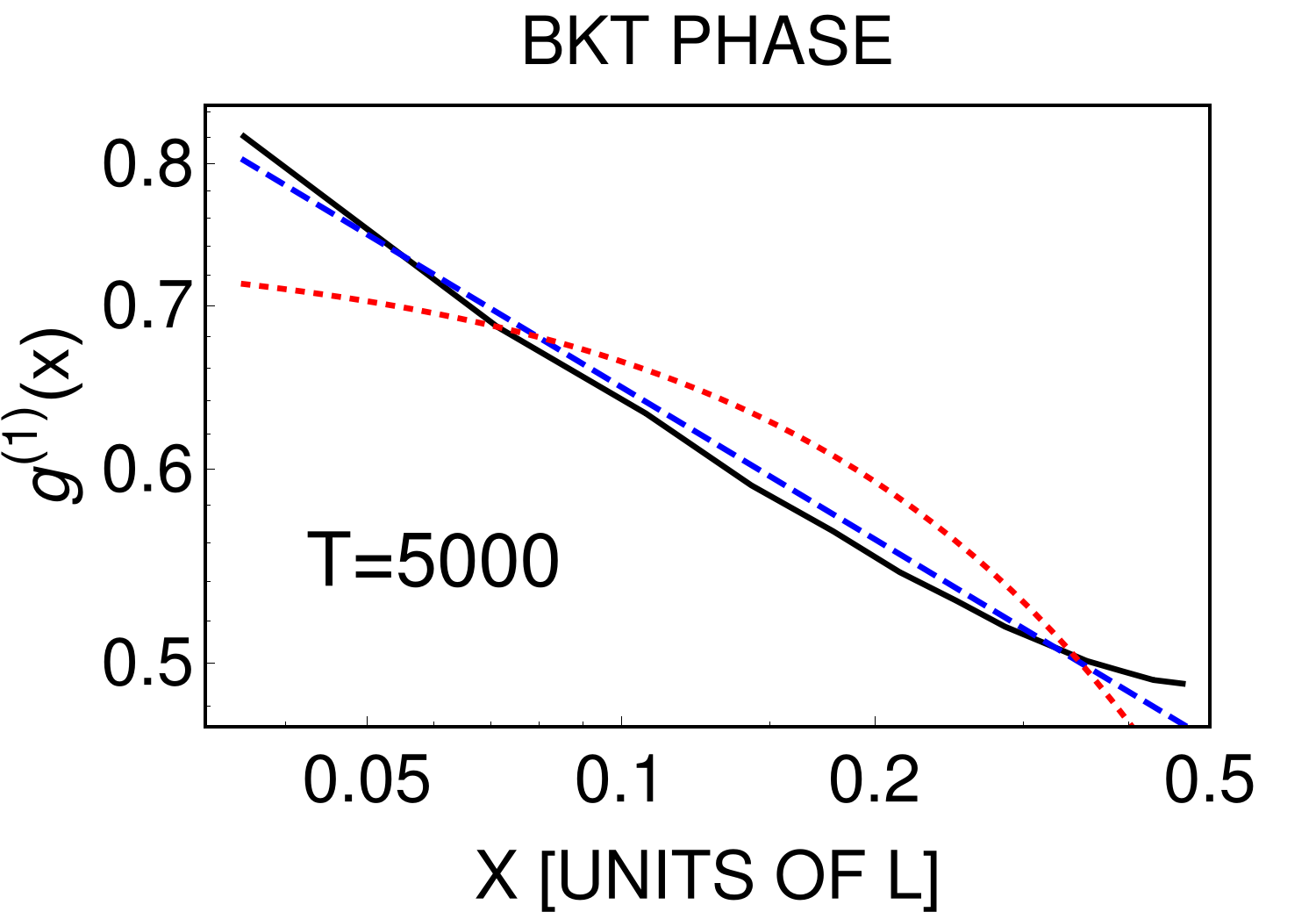} 
\includegraphics[width=5.5cm]{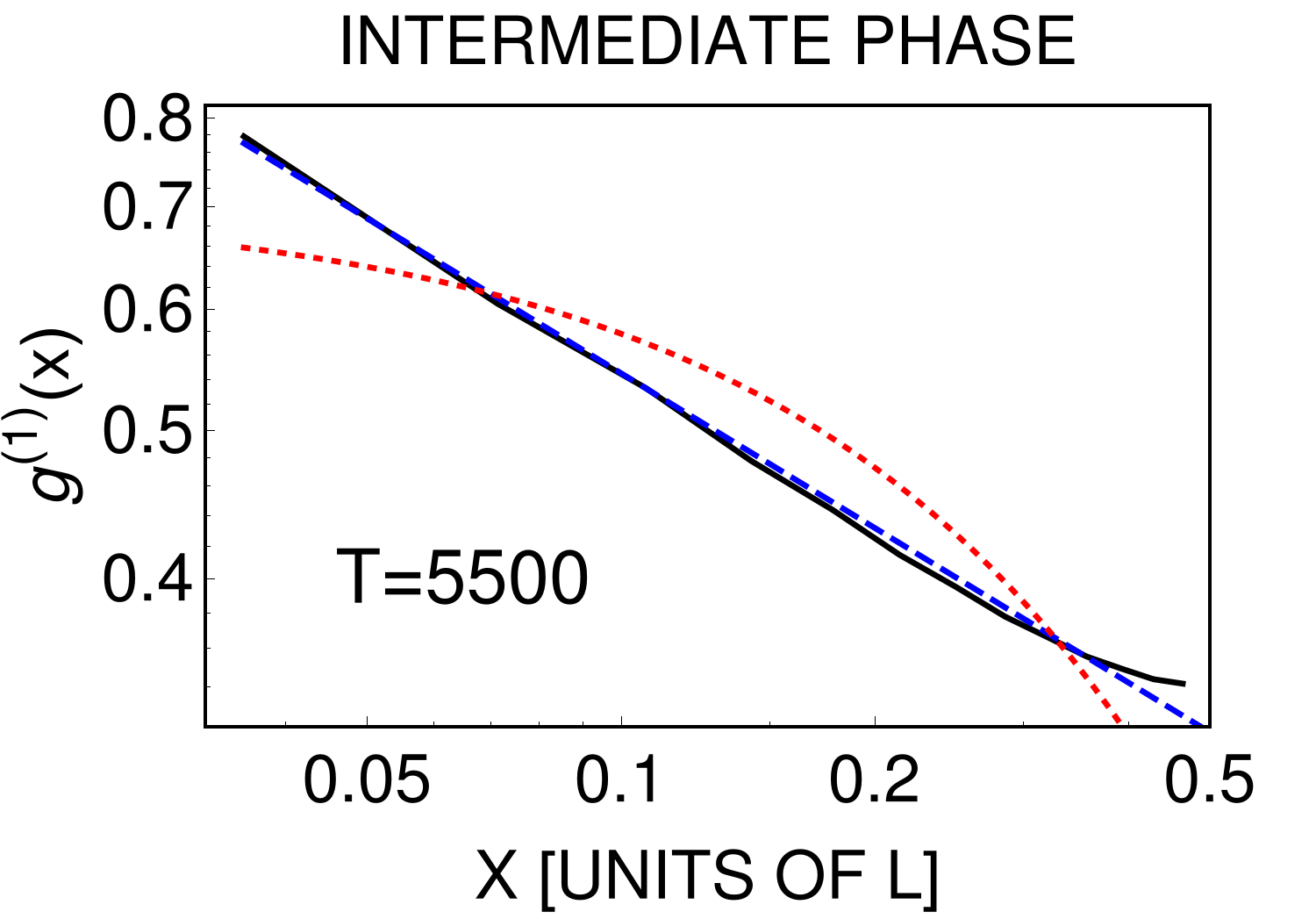}  
\includegraphics[width=5.5cm]{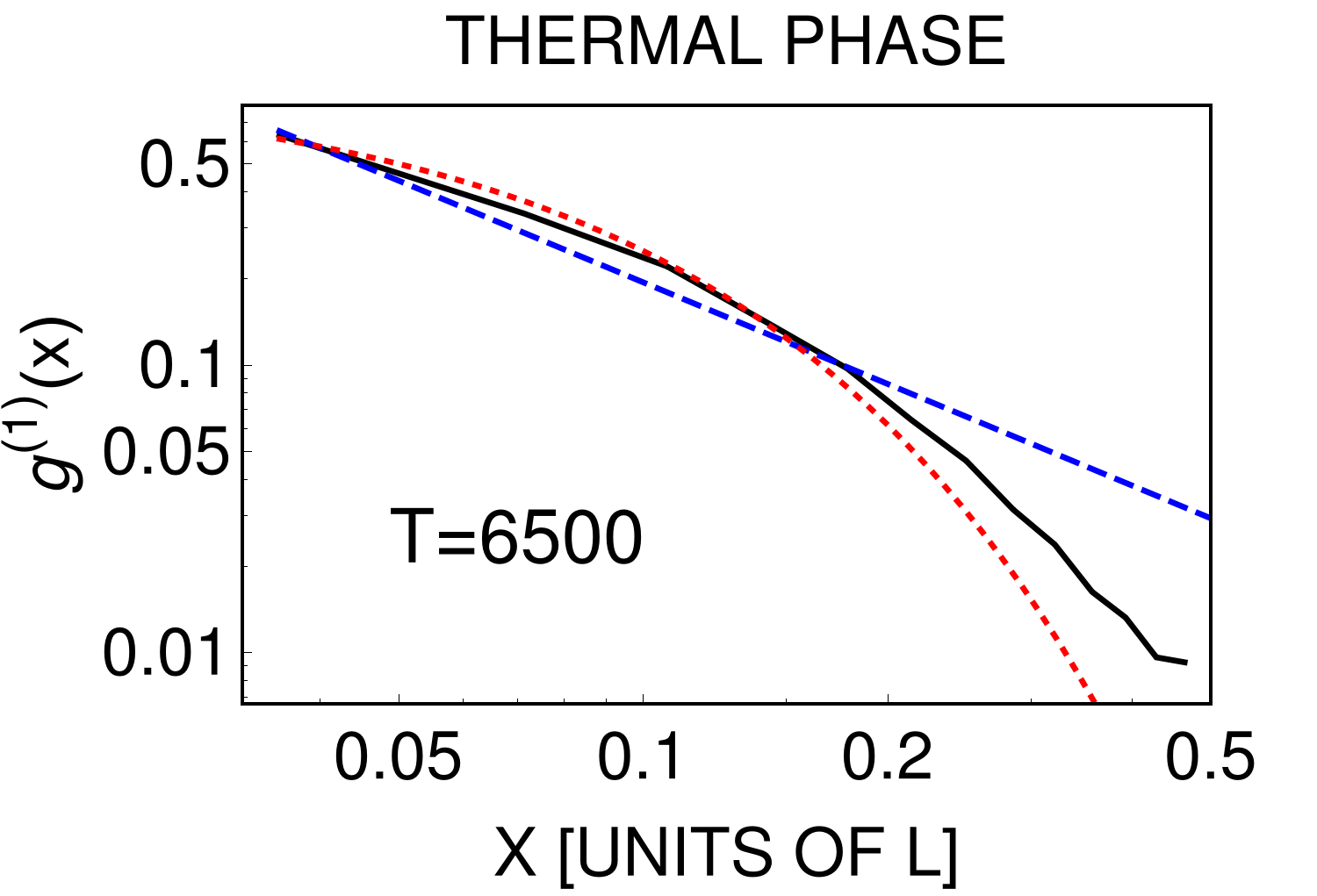}
\caption{First-order correlation function, $g^{(1)}(x)$, as in Fig. \ref{g1} but on a log-log plot.}  
\label{g1loglog}
\end{figure*}

\begin{table*}[h!tb]
\begin{tabular}{c|ccccccc}
T [box u.] & $\eta$    & $\Delta \eta$ & $\chi^2_{BKT}$ & $\beta$ &
$\Delta \beta$ & $\chi^2_{th}$ & $\chi^2_{BKT}/\chi^2_{th}$\\
\hline
5000       & $0.20702$  & $0.00650$   & $0.00138$ & $0.88434$ & $0.13184$ & $0.02257$ &  $0.06109$\\
5500       & $0.33589$  & $0.00533$   & $0.00064$ & $0.49671$ & $0.00652$ & $0.03172$ &  $0.02044$\\
6500       & $1.17454$  & $0.06814$   & $0.00900$ & $0.07182$ & $0.00311$ & $0.00281$ &  $3.20850$
\end{tabular}
\caption{Details of the fitting procedure for the BKT $(T=5000)$, intermediate $(T=5500)$, and thermal phases $(T=6500)$. $\chi^2_{BKT}$ and $\chi^2_{th}$ are calculated assuming the algebraic (dashed blue lines in Figs. \ref{g1} and \ref{g1loglog}) and the exponential (dotted red lines in Fig. \ref{g1} and \ref{g1loglog}) fits, respectively. $\Delta \eta$ shows the standard error for the exponent $\eta$, whereas $\Delta \beta$ is the standard error for the coefficient $\beta$ in the exponential decay of the correlation function, $g^{(1)} \propto \exp{(-\beta r)}$. The last column gives the ratio $\chi^2_{BKT}/\chi^2_{th}$.  }
\label{table1}
\end{table*}

In this Appendix we discuss the quality of our fitting procedure used to distinguish between different phases (BKT, intermediate, and thermal phases) of two-dimensional Bose gas, based on the calculation of the first-order correlation function (see Fig. \ref{g1}, main frames). In Fig. \ref{g1loglog} we show the correlations, $g^{(1)}(\mbf{r},0)$, in the log-log scale. Any decay according to power law changes into the linear dependence after the log-log scaling. Numerical results coming out of the CFA approximation are plotted as black solid lines. Blue dashed and red dotted lines represent the best algebraic and exponential decay fits, respectively. Fig. \ref{g1loglog}, left and middle frames, clearly demonstrate that the first-order correlation function decays algebraically both in the BKT and intermediate phases. Contrary, the right frame indicates the exponential decay of correlations in the thermal phase. Quantitative details of fitting procedure in the case of both the algebraic and exponential fits, including the $\chi^2$ test numbers, can be found in Table \ref{table1}. The $\chi^2_{BKT}$ column proves that $\chi^2$ numbers for two lowest temperatures are much smaller than for the highest temperature. It means that for two lower temperatures the algebraic fits work much better. Similarly, based on the $\chi^2_{th}$ column, one deduces that for the highest temperature, $T=6500$, the exponential fit is the correct one. The most important, however, is the last column. The ratio $\chi^2_{BKT}/\chi^2_{th}$ is very small for temperatures $T=5000$ and $T=5500$ and it becomes large for the temperature $T=6500$. It unambiguously proves that for the BKT and intermediate phases the correlations decay algebraically whereas for the thermal phase the fall off of the correlations is the exponential one.

\end{document}